\documentstyle[psfig]{cupconf}

\ifoldfss
\else
  \ifnfssone
    \newmathalphabet{\mathit}
      \addtoversion{normal}{\mathit}{cmr}{m}{it}
      \addtoversion{bold}{\mathit}{cmr}{bx}{it}
    \newmathalphabet{\mathcal}
      \addtoversion{normal}{\mathcal}{cmsy}{m}{n}
    \else
    \ifnfsstwo
    \fi
  \fi
\fi

\newcommand{\beq}{\begin{equation}}
\newcommand{\eeq}{\end{equation}}
\newcommand{\beqa}{\begin{eqnarray}}
\newcommand{\eeqa}{\end{eqnarray}}

\newcommand{\msun}{M_{\odot}}
\newcommand{\feh}{\hbox{$[\rm Fe/H]$}}

\newcommand{\dv}{\hbox{$\Delta {\rm V} ({\rm TO} - {\rm HB})$}~}

\def\hexnumber#1{\ifcase#1 0\or1\or2\or3\or4\or5\or6\or7\or8\or9\or
 A\or B\or C\or D\or E\or F\fi }

\makeatletter
\ifx\CUP@mtlplain@loaded\undefined
\else
\fi
\makeatother
 \makeatletter
 \ifx\CUP@mtlplain@loaded\undefined
   \font\tenbmi=cmmib10 at 10pt
   \font\sevenbmi=cmmib10 at 7pt
   \font\fivebmi=cmmib10 at 5pt

   \newfam\bmifam
   \textfont\bmifam=\tenbmi
   \scriptfont\bmifam=\sevenbmi
   \scriptscriptfont\bmifam=\fivebmi
   
 \fi
 \makeatother
\ifnfsstwo

\fi
\ifnfssone

\fi
\ifoldfss

\fi

\mathchardef\varLambda="0103

\makeatletter
\ifx\CUP@mtlplain@loaded\undefined
\else
\fi
\makeatother
\makeatletter
\ifx\CUP@mtlplain@loaded\undefined
  \font\tenbms=cmbsy10
  \font\sevenbms=cmbsy10 at 7pt
  \font\fivebms=cmbsy10 at 5pt
  \newfam\bmsfam
  \textfont\bmsfam=\tenbms
  \scriptfont\bmsfam=\sevenbms
  \scriptscriptfont\bmsfam=\fivebms

  \edef\bsy@{\hexnumber\bmsfam}
  \mathchardef\bnabla="0\bsy@72
\fi
\makeatother
\def\etal{\mbox{\it et al.}}

\title[Globular Cluster Ages]{Ages of Galactic Globular Clusters from
the New Yale Isochrones}

\author[Brian Chaboyer {\it et al.\/}]%
{B\ls r\ls i\ls a\ls n\ns C\ls h\ls a\ls b\ls o\ls y\ls e\ls r$^1$,\ns
P.\ns D\ls e\ls m\ls a\ls r\ls q\ls u\ls e$^2$,\ns
D.\ns B.\ns G\ls u\ls e\ls n\ls t\ls h\ls e\ls r$^3$,\ns \\
M.\ns H.\ns P\ls i\ls  n\ls s\ls o\ls n\ls n\ls e\ls a\ls u\ls l\ls t$^4$
\and \ns
L.\ns L.\ns P\ls i\ls  n\ls s\ls o\ls n\ls n\ls e\ls a\ls u\ls l\ls t$^5$
}

\affiliation{$^1$Canadian Institute for Theoretical Astrophysics,
Toronto, Ontario, Canada M5S 1A7
\\[\affilskip]
$^2$Department of Astronomy, Yale U. P.O. Box 208101, New
Haven, CT, USA 06520--8101
\\[\affilskip]
$^3$Department of Astronomy and Physics, Saint Mary's U.
Halifax, NS, Canada B3H 3C3
\\[\affilskip]
$^4$Deparment of Astronomy, Ohio State U. 174 W. 18th Ave.
Columbus, OH, USA 43210--1106
\\[\affilskip]
$^5$Houston, Texas\\
\vspace*{-6cm}to appear in {\it The Formation of the Milky Way\/}
eds. E.J. Alfaro \& G. Tenorio-Tagle\\
\vspace*{6.0cm}
}

\begin{document}
\ifnfssone
\else
  \ifnfsstwo
  \else
    \ifoldfss
      \let\mathcal\cal
      \let\mathrm\rm
      \let\mathsf\sf
    \fi
  \fi
\fi

\maketitle

\begin{abstract}
A new grid of theoretical isochrones based on the Yale stellar
evolution code using the OPAL and Kurucz opacities has been
constructed.  The grid of isochrones spans a wide range of
metallicities, helium abundances  and masses.
The construction of the isochrones is described
and the isochrones are compared to galactic globular cluster
observations.  A solar calibrated mixing length ($\alpha = 1.7$)
yields a good fit to globular cluster colour-magnitude diagrams.  Ages
for 40 globular clusters are determined using the \dv method
and the formation of the halo is discussed.
\end{abstract}

\firstsection 

\section{Introduction}
There have been important recent advances in the calculation of
stellar opacities (Iglesias \& Rogers 1991; Kurucz 1991, and private
communication).  For this reason, we have
embarked on a project to construct a new grid a theoretical isochrones
which use the latest avaliable physics, and cover a wide range of
physical parameters.  The scientific goals of this project are
three-fold: (i) test the theory of stellar structure and evolution by
using state of the art physics to construct isochrones and compare to
observations; (ii) provide a uniform grid of isochrones which can be
used to determine stellar ages; and (iii) upgrade the building blocks
for population synthesis to study distant galaxies and galaxy
formation. In my talk, I will only discuss the first two points.

\section{Isochrone Construction}
New stellar evolution models have been calculated which extend from
the zero-age main sequence to the tip of the giant branch (the
construction of core $^4$He burning stars is still in progress).  These
models use the OPAL high temperature opacities (Iglesias \& Rogers
1991) and the low temperature opacities are from Kurucz (1991).  The
nuclear reaction rates are from Bahcall \& Pinsonneault (1992).
Stellar models with masses ranging from $0.5~\msun$ to $7.0~\msun$
were calculated.  The heavy element composition of the models varied
from $Z=2\times 10^{-6}$ to $Z=0.10$. For each metallicity, two to four
different $^4$He abundances were used.  One of the great uncertainties
in present stellar evolution calculations is the treatment of
convection.  For this reason, we have varied two parameters which deal
with convection: (i) the mixing length, and (ii) the amount of
over-mixing beyond the convective core.

The conversion from stellar models to isochrones was performed using
the method of equal evolutionary points (Prather 1976).  The
construction of the isochrones themselves is independent of the choice
of colour calibration.  In this paper, the semi-empirical colour
calibration of Green \etal ~(1987) was used to transform from the
theoretical luminosities and temperatures to observational magnitudes
and colours.  The uncertainties are largest at the lowest
temperatures.  The stellar models use a scaled solar heavy element
mixture.  The effects of $\alpha$-element enhancement are taken into
account by adjusting the relationship between \feh ~and total $Z$
(Salaris \etal ~1993).  The full isochrone grid is tabulated in Table
\ref{grid}.  These isochrones will be available via anonymous ftp
(relase date $\sim$ Janurary, 1995),
contact {\tt demarque@astro.yale.edu} for further details.

\begin{table}
  \begin{center}
  \begin{tabular}{ccccccl}
                  &&&                 & Mixing & over-shoot      \\
      $Z$& \feh&$[\alpha/{\rm Fe}]$&$Y$ &Length  &  (H$_p$) & Age (Gyr)\\[3pt]
$2\times 10^{-6}$ &$-4.3$&0.4&0.20, 0.23, 0.26 & 1.5, 1.7, 2.0 & 0.0&4 -- 22 \\
$2\times 10^{-5}$ &$-3.3$&0.4&0.20, 0.23, 0.26 & 1.5, 1.7, 2.0 & 0.0&4 -- 22 \\
$6\times 10^{-5}$ &$-2.8$&0.4&0.20, 0.23, 0.26 & 1.5, 1.7, 2.0 & 0.0&4 -- 22 \\
$2\times 10^{-4}$ &$-2.3$&0.4&0.20, 0.23, 0.26 & 1.5, 1.7, 2.0 & 0.0&2 -- 22 \\
$6\times 10^{-4}$ &$-1.8$&0.4&0.20, 0.23, 0.26 & 1.5, 1.7, 2.0 & 0.0&2 -- 22\\
$2\times 10^{-4}$ &$-1.3$&0.4&0.20, 0.23, 0.26 & 1.5, 1.7, 2.0 & 0.0&2 -- 22 \\
$4\times 10^{-4}$ &$-0.9$&0.3&0.20, 0.23, 0.26 & 1.5, 1.7, 2.0 & 0.0 &0.5 --
22\\
$7\times 10^{-4}$ &$-0.6$&0.2&0.20, 0.23, 0.26 & 1.5, 1.7, 2.0 & 0.0 &0.5 --
22\\
$0.01 $          &$-0.4$&0.2&0.25, 0.27, 0.30, 0.35 & 1.7 & 0.0, 0.1 &0.05 --
22\\
$0.02 $          &$+0.0$&0.0&0.25, 0.27, 0.30, 0.35 & 1.5, 1.7, 2.0 & 0.0, 0.1
&0.05 -- 22\\
$0.04 $          &$+0.3$&0.0&0.32,  0.34 & 1.7 & 0.0 & 0.05 -- 22 \\
$0.06 $          &$+0.5$&0.0&0.36,  0.40 & 1.7 & 0.0 & 0.05 -- 22 \\
$0.10 $          &$+0.7$&0.0&0.44,  0.52 & 1.7 & 0.0 & 0.05 -- 22 \\
  \end{tabular}
  \caption{Parameters for our grid of isochrones.}
\label{grid}
\end{center}
\end{table}

\section{Isochrone Fits}
The isochrones have been compared to observed colour magnitude
diagrams (CMDs) for a variety of globular clusters.  A typical example
is shown in Figure \ref{cmd}, which shows the fit to M68 ($\feh = -2.1$)
for two different mixing lengths: $\alpha = 1.7$ and $\alpha = 2.0$.
The 16 Gyr $\alpha = 1.7$ isochrone provides a superb fit from the
lower main sequence, through the main sequence turn-off and sub-giant
branch, all the way to the tip of the red giant branch. The $\alpha =
2.0$ (and $\alpha = 1.5$) isochrones do not provide a good match
to the observations for
any reasonable values of the reddening and distance modulus.
Similar conclusions are reached for globular clusters with higher
metallicities, though our fit to shape of the 47 Tuc ($\feh = -0.7$)
CMD is not as good.  Our solar
solar calibrated models require a mixing length
of $\alpha = 1.7$.  Thus, one value of the mixing length provides a good
match to the sun and metal-poor stars.
\begin{figure}
\vspace*{10.0cm}
\caption{The observations of M68 (dots)
by Walker (1994) are compared to the
best fitting theoretical isochrones with ages of 14, 16 and 18 Gyr
(solid lines).  Isochrones with a mixing length of 1.7 are shown in
the left panel, while those with $\alpha = 2.0$ are shown in the right
panel.  The reddening and distance modulus used to shift the
isochrones are shown on the graph.}
\label{cmd}
\end{figure}

\section{\dv Ages}
A commonly used age indicator which is insensitive to reddening and
uncertainties in our treatment of convection and stellar atmospheres
is the difference in magnitude between the turn-off and the horizontal
branch, \dv (e.g.~Iben \& Renzini 1984).  The ages
derived using \dv are well understood theoretically as they do not
depend on the model colours.  As we have not yet calculated horizontal
branch models, the turn-off
magnitudes from our isochrones have been combined with the observed RR
Lyr magnitudes (${\rm M_v} = 0.20\feh + 1.06$, Carney \etal ~1992;
Skillen \etal ~1993) to calculate ages for 40 halo globular clusters
which have observed \dv values.  An age
spread of $\sim 4$ Gyr is clearly present.  There is no significant
correlation between age and galactocentric distance, but an
age-metallicity relationship does exist in the outer halo, and is
shown in Figure
\ref{agefeh}. From this, we can conclude that the halo of our galaxy
formed over an extended period of time, with an gradual enrichment in
the heavy element content in the outer halo.
\begin{figure}
\vspace*{8.0cm}
\caption{Ages of globular clusters as a function of metallicity for
the  inner halo (upper panel) and outer halo (lower panel).  The solid
line shows a least squares fit to the data. The correlation is
significant at greater than the 99\% confidence level for the outer
halo clusters.
In contrast, the significance of the correlation for the inner halo
clusters is only 92\%. }
\label{agefeh}
\end{figure}

\begin{acknowledgments}
BC is supported by  NSERC of Canada.
Research supported in part by NASA grants
NAG5--1486, NAGW--2136, NAGW--2469 and  NAGW-2531 to Yale University.
\end{acknowledgments}

\noindent
{\large\bf Discussion}
\vspace*{4pt}

\noindent
A.~APARICIO:~~The observed luminosity function is a powerful tool to test the
goodness of stellar evolutionary models.  How good is the agreement
with your models?

\noindent
B.~CHABOYER:~~Our models do a good job of predicting the relative
numbers of stars on the red giant branch.  At present, there are few
observed luminosity functions which extend from the main sequence up
the giant branch.  The only one I am aware of was recently published
by M.~Bolte.  There appears to be a discrepancy between the theory
and observations.  More observations are needed to access the
significance of this problem.

\noindent
A.~APARICIO:~~What is the critical mass for the helium shell flash in
your models?

\noindent
B.~CHABOYER:~~I have not examined this issue.

\noindent
G.~CARRARO:~~A point about the Green \etal (1987) colour transformation.
Basically they were built up to study the metal poor globular
clusters  population.  At what level of confidence can they be used
for supersolar, younger clusters  (e.g.~NGC 6791)?

\noindent
B.~CHABOYER:~~I have not looked into this question too closely, as I
have only been working on the metal poor part of this project.

\noindent
J.~KALUZNY:~~It seems to me that your isochrones do not provide a good
fit for the lower main-sequence.  Some systematic deviations are visible
on the $M_V$ vs. $(B-V)$ and $M_V$ vs. $(V-I)$ CMDs.

\noindent
B.~CHABOYER:~~The CMDs I have compared my isochrones to  were picked
because they
included extensive giant branchs.  These CMDs are not that deep, so
that there is a fair bit of scatter in the observations on the lower
main sequence.  I don't see any large systematic deviations between
the theory and observations on the lower main sequence (see figure
\ref{cmd}).  It would be
best to compare the isochrones to some deep CMDs in order to fully
address this question.
\end{document}